\documentclass{article}

\usepackage{PRIMEarxiv}

\usepackage[utf8]{inputenc} 
\usepackage[T1]{fontenc}    
\usepackage{hyperref}       
\usepackage{url}            
\usepackage{booktabs}       
\usepackage{amsfonts}       
\usepackage{amsmath}
\usepackage{amssymb}
\usepackage{nicefrac}       
\usepackage{microtype}      
\usepackage{lipsum}
\usepackage{graphicx}
\usepackage{subfigure}
\usepackage{mathpazo}
\usepackage{sectsty}
\usepackage{tikz}

\usepackage{booktabs}
\usepackage{subcaption}
\usepackage{multirow}
\usepackage{array}
\usepackage{color}

\usepackage{pifont}
\usepackage{graphicx}
\usepackage{caption}
\usepackage[authoryear]{natbib}
\usepackage{lineno,hyperref}
\usepackage{subfigure}
\usepackage{color}

\usepackage{amsmath,amssymb,bm}
\usepackage{booktabs,multirow,array,tabularx,colortbl,longtable}
\usepackage{siunitx}
\usepackage{expl3}
\usepackage{tikz}
\usetikzlibrary{arrows.meta,positioning,calc,fit,shapes.geometric}
\usepackage{pgfplots}
\pgfplotsset{compat=1.17}
\usepackage{float}
\usepackage{microtype}
\usepackage{enumitem}
\usepackage{url}

\graphicspath{{media/}}     

\newcommand{\gradS}{\nabla_{\!s}}
\newcommand{\Ssurf}{\mathcal{S}}

\newcommand{\Lref}{L_{\mathrm{ref}}}

\title{Geometry-Based Metrics for Early-Stage Hull-Form Producibility Screening
}

\author{
  A. Serani$^{1,\star}$ and K. Maki$^2$\\
  $^1$National Research Council-Institute of Marine Engineering, Rome, Italy\\
  $^2$University of Michigan, Ann Arbor, USA\\
  $^\star$\texttt{andrea.serani@cnr.it} \\
}

\begin{document}

\begin{tikzpicture}[remember picture,overlay]
   \node [rectangle, fill=cyan, fill opacity=0.5, anchor=north, minimum width=\paperwidth, minimum height=3cm] at (current page.north) {};

   \node [anchor=north, minimum width=\paperwidth, minimum height=3cm, text width=\textwidth, align=center, text height=5ex, text depth=15ex, align=left] at (current page.north) {
     \sffamily\small
     \textbf{This is a preprint submitted to:} \textit{Ocean Engineering}
   };
\end{tikzpicture}

\maketitle

\begin{abstract}
This paper presents a representation-aware framework for geometry-based screening of hull-form producibility at early design stages. The proposed signature combines dimensionless total and signed developability deviation with curvature-class area fractions, distributed fields, metric-specific validity, and representation provenance. These descriptors characterize surface features relevant to plate forming and developability, but are not calibrated predictors of fabrication cost, forming effort, or process feasibility. Native IGES/STEP boundary representations (BReps) are evaluated through direct differential geometry and trimmed-domain quadrature, whereas triangulated surfaces use discrete curvature recovery and area-weighted aggregation. Analytical and semi-analytical controls verify the formulation, while matched-face BRep-to-mesh tests assess discrete curvature recovery. Application to DTMB 5415, KCS, JBC, and KVLCC2M shows that curvature intensity and areal extent provide complementary information and that derivative-based outcomes can be representation sensitive. KCS, for example, exhibits approximately 24\% greater developability deviation than DTMB 5415, while double-curved regions occupy 72.9\% of its valid surface versus nearly the entire DTMB valid surface. The resulting quantities provide an early geometric screening layer for subsequent use as objectives, constraints, surrogate responses, or design-space features. \texttt{HullProd}, the companion open-source software, implements the signature, distributed fields, validity, and provenance.
\end{abstract}


\section{Introduction}
Ship hull-form design is increasingly supported by simulation-based design optimization, surrogate modeling, and automated design-space exploration \citep{serani2024scoping}. These methods enable large families of geometries to be compared with respect to resistance, powering, seakeeping, maneuvering, stability, and mission requirements \citep{Zhu2024}. Nevertheless, a geometry that is attractive from a hydrodynamic viewpoint is not necessarily attractive from a production viewpoint. Parametric deformation, dimensionality-reduced shape representations, and generative geometry methods may produce extensive double curvature, local curvature concentration, or abrupt changes in surface shape that are difficult to identify if producibility is considered only after a design has been selected.

The fabrication of curved shell plates has long motivated research on plate classification and geometric producibility \citep{Letcher1993,Kim2004,Kim2006,Kim2016}; plate-deformation mechanics and forming analysis \citep{ShinRyu2000}; line-heating and thermal-forming methods \citep{Jang1997,JangMoon1998,Shin2003,Shin2004,Park2018}; springback compensation and plate flattening \citep{Hwang2010,Zhang2019}; localization and fabrication support \citep{Park2007,Park2008}; and dimensional inspection and fabrication-completeness assessment \citep{Wang2016,Kim2024}. Early design-for-production studies identified double-curvature plating as a source of additional forming effort \citep{KraineIngvason1990}, emphasized integrating design and manufacture before production-sensitive choices become fixed \citep{EcclesMarcus1992}, and documented the need for standardized quantitative measures and tools for early-stage producibility trade-offs \citep{KeaneFireman1992}.
These studies show that production difficulty depends on the target geometry together with plate size, material, thickness, forming route, tolerances, equipment, and shipyard practice. Geometry alone therefore cannot predict fabrication man-hours or cost. It can, however, expose features known to influence plate forming, fairing, subdivision, and inspection before detailed production information is available.

\cite{BrownBarentine1996} already used Gaussian surface curvature to identify producibility-relevant regions of a naval combatant hull and to examine performance, cost, and producibility trade-offs during concept design. \cite{Parsons1999} proposed a scalar hull-form producibility metric based on area-weighted curvature classes and relative fabrication-cost coefficients, establishing an early quantitative connection between hull geometry and production considerations. Subsequent work extended geometry-based manufacturing-cost estimation to complex offshore structures \citep{Nam2012} and incorporated producibility metrics into multi-objective hull-form optimization \citep{TempleCollette2016}. In parallel, curvature information has been used for curved-plate classification and forming-route selection \citep{Kim2004,Kim2006,Kim2016}, forming-cycle-time estimation \citep{Song2022}, line-heating process planning \citep{JangMoon1998,Shin2003,Shin2004,Park2018}, and developability- and panelization-oriented surface design \citep{ChalfantMaekawa1998,PerezClemente2011,Gavriil2019,Takezawa2019}.
These studies demonstrate the production relevance of curvature magnitude, sign, distribution, and variation, but they address different levels of the production problem. Cost-weighted metrics require empirical or process-specific coefficients, while plate-classification and forming models rely on information such as plate size, material, thickness, forming route, tolerances, equipment, or yard practice. A remaining methodological gap is a normalized, representation-aware, and validity-qualified geometric signature that separates curvature intensity, sign, and areal composition, provides distributed localization, and can be evaluated consistently through native parametric boundary representation (BRep) and triangle-mesh backends without embedding process-specific cost weights.

The objective of the present work is to formulate, verify, and implement such a set of descriptors for geometry-induced hull-form producibility. Rather than collapsing the assessment into a single universal score, the proposed framework retains complementary information on double-curvature intensity, sign, and areal extent, together with distributed localization, metric-specific validity, and representation provenance, recording the source representation and numerical realization associated with each result. The present scope is deliberately limited to the surface geometry itself. Factors such as structural arrangement, panel subdivision, material and thickness, forming route, tolerances, equipment, shipyard practice, measured deviations, cost, and schedule all shape production feasibility in ways that geometry alone cannot reveal, and none of them fall within the present assessment. The resulting framework is therefore intended as an initial geometric screening layer through which candidate hull forms can be compared and interpreted before detailed production information becomes available.

In the present paper, a \emph{geometry-based producibility metric} is defined as a normalized and interpretable descriptor linked to a forming- or geometry-quality mechanism. The proposed \emph{geometry-based producibility signature} combines, for each metric family, a global descriptor, a distributed surface field, and a metric-specific validity and provenance record. Global quantities enable compact cross-hull comparison, distributed outputs reveal where each response originates, and the validity record shows whether a reported quantity is finite, conditional, unresolved, or inapplicable. Subsequent use of these descriptors in optimization, surrogate modeling, or design-space dimensionality reduction is considered only as a downstream application.

To make the metrics applicable to the geometric representations available during design, the geometric functional is separated from its numerical realization. A hull surface may be supplied either as a native BRep or as a triangulated surface mesh, and the corresponding evaluation is formulated to remain consistent with the geometry represented by each description. For triangulated surfaces, curvature estimation is inherently affected by mesh resolution, connectivity, boundary treatment, and local element quality \citep{Meyer2003,Wardetzky2007}; these effects are therefore treated as part of the numerical realization rather than of the metric definition itself. Because the proposed descriptors impose different differential and integral requirements, validity is assessed individually for each metric and reported together with the representation provenance.

The paper makes three principal contributions. First, it formulates a normalized, interpretable geometry-based signature combining total developability intensity, signed elliptic and saddle/reverse contributions, curvature-class areal composition, and distributed localization. Second, it establishes a representation-aware numerical contract that separates native BRep and triangle-mesh realizations and accompanies each result with metric-specific validity, convergence or regularity evidence, and representation provenance. Third, analytical and semi-analytical controls and four public hulls assess the continuous formulation, its numerical realizations, and representation sensitivity, with the complete retained contract implemented in the companion \texttt{HullProd} software. These contributions are situated within the broader ship-production literature without claiming to cover the full scope of the production problem. Additional higher-regularity and section-based quantities were screened during development but were not promoted to the recommended signature after numerical verification.

The remainder of the paper is organized as follows. Section~\ref{sec:background} reviews the relation to ship-production metrics. Section~\ref{sec:framework} defines the continuous descriptors. Section~\ref{sec:numerics} presents the two numerical backends and validity contract. Section~\ref{sec:verification} summarizes the analytical and public-hull protocol. Section~\ref{sec:results} proceeds from analytical verification to global public-hull signatures, distributed localization, the validity result, and representation sensitivity. Section~\ref{sec:discussion} synthesizes the engineering hierarchy and principal validation gap. Conclusions are given in Section~\ref{sec:conclusions}.

\section{Background and relation to ship-production metrics}
\label{sec:background}

Early-stage producibility studies seek information that can shape the hull form while design choices are still open, before structural and production decisions become difficult to revise. Geometry is attractive because it is available early, can be evaluated consistently across design variants, and can be introduced as an objective or constraint in automated design. Its limitation is equally important: the same surface can imply different effort under different materials, plate subdivisions, forming processes, tolerances, and facilities. Geometry-based descriptors should therefore be interpreted as indicators of shape-induced complexity, not as direct work-content estimators.

\begin{figure*}[!t]
\centering
\includegraphics[width=\textwidth]{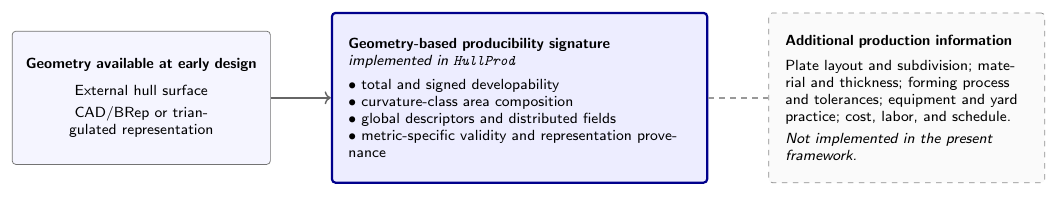}
\caption{Positioning of the geometry-based producibility signature within early-stage hull design. The retained signature, implemented in \texttt{HullProd}, is evaluated from the external hull surface and combines global and distributed descriptors with metric-specific validity and representation provenance. Plate layout, material, forming process, yard practice, cost, labor, and schedule require additional information and are not modeled in the present framework.}
\label{fig:scope}
\end{figure*}

\subsection{Developability, curvature class, and plate forming}
Developable surfaces have zero Gaussian curvature and can, ideally, be formed from a planar sheet without in-plane stretching. Cylindrical and conical surfaces are the canonical examples. Elliptic and saddle surfaces have nonzero Gaussian curvature and require in-plane strain, localized heating, stretching, shrinking, or subdivision into smaller plates. Earlier ship-design-for-production work explicitly identified double-curvature hull plating as a source of increased forming effort \citep{KraineIngvason1990}, and \cite{BrownBarentine1996} subsequently used Gaussian curvature directly to distinguish more- and less-producible regions of a naval combatant hull form. The present formulation does not introduce Gaussian curvature as a new producibility concept; it formalizes this geometric lineage into normalized total and signed surface descriptors, areal curvature composition, distributed localization, and representation-aware validity and provenance. \cite{Letcher1993} and subsequent plate-forming work provide the physical basis for using Gaussian curvature and principal-curvature signs as forming-relevant geometric descriptors.

\cite{Parsons1999} converted curvature classes into a scalar hull measure using relative fabrication weights. The present work retains the geometry-only class fractions but does not adopt universal cost coefficients. Kim et al. \citep{Kim2004,Kim2006,Kim2016,Song2022} use curvature information within process-specific classification and cycle-time models. Those approaches remain essential once plate, material, process, and yard data are available; they are complementary rather than competing with the present surface-only assessment.

\subsection{Scope relative to cost and process models}

The proposed framework occupies an intentionally limited layer of the broader
producibility problem. It uses the external hull surface to quantify
geometry-induced complexity before plate subdivision, material, thickness,
forming process, tolerances, equipment, and shipyard practice are specified
(Fig.~\ref{fig:scope}). These additional factors ultimately
determine fabrication effort, cost, labor, and schedule, but cannot be inferred
from geometry alone.

The retained geometric layer is implemented in \texttt{HullProd}, which
evaluates the recommended signature from native BRep or triangulated hull
representations and reports global and distributed quantities together with
metric-specific validity and representation provenance. No process-specific
weights or calibrated shipyard cost model are introduced.

Table~\ref{tab:literatureposition} describes the proposed descriptors
relative to the main literature streams. Rather than providing a taxonomy of
the complete production problem, it distinguishes established capabilities
from the elements retained and extended in the present framework.
Table~\ref{tab:metricroles} then summarizes the intended engineering role
of each retained signature family before the numerical results are considered.

\subsection{Screened descriptors and representation diagnostics}
Hull-fairing and surface-quality research also motivated the investigation of higher-regularity surface and section-based descriptors. These quantities were screened during framework development but are not part of the recommended signature because the numerical assessment did not establish robust general hull-level values across the considered public CAD representations.

Mesh-local normal variation is likewise retained only as a
representation-quality diagnostic because it depends directly on mesh
resolution, connectivity, and local element quality. The corresponding
numerical evidence for the screened quantities and representation-sensitive
diagnostics is discussed in Sections~\ref{sec:results} and
\ref{sec:discussion}.

\begin{table*}[!t]
\centering
\caption{Positioning of the present geometry-based producibility signature relative to previous work.}
\label{tab:literatureposition}
\scriptsize
\renewcommand{\arraystretch}{1.10}
\setlength{\tabcolsep}{3.0pt}
\begin{tabularx}{\textwidth}{p{0.14\textwidth}p{0.21\textwidth}p{0.25\textwidth}X}
\toprule
Research stream & Representative references & Established capability & Role in the present work \\
\midrule
Hull fairing and surface quality & Sariöz \cite{Sarioz2006} & Curvature smoothness and fairing objectives for hull lines and surfaces. & Motivates screened geometry-regularity diagnostics; higher-regularity quantities are not retained in the recommended signature. \\
Developability and geometric paneling & Letcher \cite{Letcher1993}; Chalfant and Maekawa \cite{ChalfantMaekawa1998}; Pérez and Clemente \cite{PerezClemente2011}; Gavriil et al. \cite{Gavriil2019} & Developable or quasi-developable surface design, Gaussian-curvature interpretation, and panel approximation. & Supports continuous developability density, signed elliptic/saddle contributions, and the global deviation from $K=0$. \\
Curved-plate classification and forming & Kraine and Ingvason \cite{KraineIngvason1990}; Jang et al. \cite{Jang1997}; Jang and Moon \cite{JangMoon1998}; Shin and Ryu \cite{ShinRyu2000}; Shin et al. \cite{Shin2003,Shin2004}; Park et al. \cite{Park2007,Park2008}; Kim et al. \cite{Kim2004,Kim2006,Kim2016}; Song et al. \cite{Song2022} & Double-curvature forming guidance, curvature-based plate classes, forming-route and line-heating planning, and process-specific predictors. & Supports geometric curvature-class maps and the distinction between elliptic and saddle/reverse double curvature, without assigning universal process costs. \\
Quantitative producibility, cost proxies, and optimization & Keane and Fireman \cite{KeaneFireman1992}; Brown and Barentine \cite{BrownBarentine1996}; Parsons et al. \cite{Parsons1999}; Nam et al. \cite{Nam2012}; Temple and Collette \cite{TempleCollette2016} & Quantitative-tool needs, Gaussian-curvature identification of producibility-relevant hull regions, relative-cost indices, and concept-design trade-off studies. & Establishes early-stage precedents; the present work adds normalized intensity, sign, and extent, distributed fields, and representation-aware validity and provenance without universal cost weights. \\
Localization, subdivision, and inspection & Park et al. \cite{Park2007,Park2008}; Kim et al. \cite{Kim2012}; Hiekata et al. \cite{Hiekata2011}; Wang et al. \cite{Wang2016}; Kim et al. \cite{Kim2024} & Plate subdivision, fabricated-plate localization, measured deviations, and inspection workflows. & Identifies important production information that requires plate, process, or measurement data and therefore lies outside the external-surface-only assessment. \\
Computational curvature on triangulated surfaces & Meyer et al. \cite{Meyer2003}; Wardetzky et al. \cite{Wardetzky2007}; Rusinkiewicz \cite{Rusinkiewicz2004}; Cazals and Pouget \cite{CazalsPouget2005}; Lachaud et al. \cite{Lachaud2020} & Discrete curvature, local recovery, consistency limits, and corrected curvature measures. & Supports the mesh backend and motivates representation-aware verification, mesh-quality reporting, and metric-specific validity. \\
\bottomrule
\end{tabularx}
\renewcommand{\arraystretch}{1.0}
\end{table*}

\begin{table*}[!t]
\centering
\caption{Engineering interpretation and current role of the retained geometry-based signature families.}
\label{tab:metricroles}
\small
\begin{tabularx}{\textwidth}{p{0.15\textwidth}X p{0.33\textwidth}}
\toprule
Quantity & Geometric meaning and connection to producibility & Current interpretation \\
\midrule
$I_D,I_D^\pm$ & Departure from developability and its sign. Nonzero $K$ implies double curvature and in-plane deformation of an initially flat plate. & Principal forming-related descriptor when the $|K|$ integral is valid. \\
$\mathbf{a}_C$ & Area composition into flat, singly curved, elliptic, and saddle regions, retaining the connection to geometric plate classes. & Descriptive surface composition; quasi-zero thresholds are not calibrated process classes. \\
\bottomrule
\end{tabularx}
\end{table*}

\section{Geometry-based producibility framework}
\label{sec:framework}

Let $\Ssurf\subset\mathbb{R}^3$ denote the represented external hull surface, with area $A$ and reference length $\Lref$. At a sufficiently regular point of $\Ssurf$, let $\kappa_1$ and $\kappa_2$ denote the principal curvatures. The corresponding mean curvature $H$ and Gaussian curvature $K$ are defined as
\begin{equation}
H=\frac{\kappa_1+\kappa_2}{2},
\qquad
K=\kappa_1\kappa_2.
\label{eq:HK}
\end{equation}
The mean curvature characterizes the local average bending of the surface, while the Gaussian curvature distinguishes locally developable regions from elliptic or saddle-type double curvature through its magnitude and sign. Vanishing Gaussian curvature over a sufficiently regular surface region is the local differential condition for developability; pointwise $K=0$ alone does not establish developability of a finite surface patch. At a regular point, $K>0$ identifies elliptic double curvature, $K<0$ saddle or reverse curvature, and $K=0$ is consistent with flat or singly curved local geometry.

The adopted Gaussian-curvature scale is
\begin{equation}
K_{\rm ref}=L_{\rm ref}^{-2}.
\label{eq:refs}
\end{equation}
The reference length $\Lref$ is a hull-level scale used to nondimensionalize
the curvature descriptors and remove trivial dimensional scaling. Any
consistently defined hull-scale length may be adopted provided it is reported;
however, quantitative cross-design comparisons require the same reference-length
convention. In the present public-hull campaign, $\Lref=L_{pp}$.
This scale is used to construct the dimensionless Gaussian-curvature descriptors. The quantities introduced below are defined on the subset of $\Ssurf$ where the corresponding differential properties exist and satisfy the required regularity conditions.

\subsection{Developability deviation}
Developability provides the most direct geometric connection between surface
curvature and the forming of an initially flat plate. A sufficiently smooth,
regular surface region is locally developable when its Gaussian curvature
vanishes throughout the region. Planes, cylinders, and cones are canonical examples: although they may
have nonzero mean curvature, they can ideally be unfolded onto a plane without
in-plane stretching. In contrast, a region with $K\neq0$ is doubly curved and
cannot be flattened isometrically. Its realization from an initially flat
plate therefore requires in-plane deformation, subdivision, or a
process-specific forming operation.

The magnitude of Gaussian curvature is therefore used here to quantify the
local departure from developability. Using the reference scale
$K_{\rm ref}=L{\rm ref}^{-2}$, the nondimensional local developability-deviation field is
\begin{equation}
q_D(\mathbf{x})
=
\frac{|K(\mathbf{x})|}{K_{\rm ref}}
=
|K(\mathbf{x})|L_{\rm ref}^2 .
\label{eq:devfield}
\end{equation}
Thus, $q_D=0$ marks zero Gaussian curvature at a sample and, when it holds over
a sufficiently regular region, the local differential condition for
developability. Increasing $q_D$ indicates increasing double-curvature
intensity relative to the hull reference length.

The corresponding hull-level descriptor is the area average over the valid
Gaussian-curvature domain,
\begin{equation}
I_D
=
\frac{1}{A_K}
\int_{\Ssurf_K} q_D(\mathbf{x})\,\mathrm dA ,
\label{eq:dev}
\end{equation}
where $\Ssurf_K\subseteq\Ssurf$ is the portion of the represented surface on
which Gaussian curvature is defined and integrable, and $A_K$ is its area.
The normalization prevents the global descriptor from scaling trivially with
the represented surface area. The local field $q_D(\mathbf{x})$ identifies where
double curvature is concentrated, whereas $I_D$ provides a compact measure of
its overall intensity.

Accordingly, $I_D=0$ is the developable limit; all else being equal, lower
$I_D$ indicates a surface closer to developability, whereas higher $I_D$
indicates greater average double-curvature intensity. This directionality is
geometric and should not be interpreted as a calibrated scale of forming
effort or fabrication cost.

The sign of Gaussian curvature contains additional geometric information that
is lost in $|K|$. Positive Gaussian curvature corresponds to elliptic
(or synclastic) double curvature, whereas negative Gaussian curvature
corresponds to saddle or reverse (anticlastic) curvature. The total field is
therefore decomposed into positive and negative contributions,
\begin{equation}
q_D^+(\mathbf{x})=\max(K(\mathbf{x}),0)\Lref^2,
\end{equation}
\begin{equation}
q_D^-(\mathbf{x})=\max(-K(\mathbf{x}),0)\Lref^2 ,
\label{eq:devsignedfields}
\end{equation}
with corresponding global descriptors
\begin{equation}
I_D^\pm
=
\frac{1}{A_K}
\int_{\Ssurf_K} q_D^\pm(\mathbf{x})\,\mathrm dA .
\label{eq:devsigned}
\end{equation}
By construction,
\begin{equation}
I_D=I_D^+ + I_D^- ,
\end{equation}
so that $I_D$ measures total double-curvature intensity, while
$I_D^+$ and $I_D^-$ distinguish how much of that intensity is associated
with elliptic and saddle/reverse curvature, respectively.

For an open represented surface, the signed Gaussian-curvature integral also
depends on the represented domain and its boundary geometry through the
Gauss--Bonnet relation \citep{doCarmo1976}. The balance between $I_D^+$ and $I_D^-$ must therefore be interpreted
with explicit surface-domain and boundary provenance, including whether the
authoritative source represents a half hull or a full hull.

\subsection{Curvature-class area distribution}
The developability-deviation descriptors quantify the intensity of double
curvature, but they do not describe how the different geometric surface types
are distributed over the hull. A complementary characterization is therefore
obtained by classifying each sufficiently regular surface point according to
the local combination of mean and Gaussian curvature.

In the ideal continuous setting, a flat region has both principal curvatures
equal to zero and therefore $H=K=0$. A singly curved region, such as a
cylindrical surface, has one vanishing principal curvature and consequently
$K=0$ while $H\neq0$. Regions with $K>0$ exhibit elliptic (synclastic) double
curvature, whereas regions with $K<0$ exhibit saddle or reverse (anticlastic)
double curvature. This classification separates the \emph{type} of local
surface geometry from the curvature intensity measured by the developability
deviation.

Exact zero curvature cannot generally be expected from numerical CAD or mesh
representations. The ideal classes are therefore identified using
scale-consistent quasi-zero tolerances. Let $h_f$ and $k_f$ denote prescribed
dimensionless tolerances for mean and Gaussian curvature, respectively. The
corresponding dimensional thresholds are
\begin{equation}
\tau_H=\frac{h_f}{\Lref},
\qquad
\tau_K=\frac{k_f}{\Lref^2},
\label{eq:class_thresholds}
\end{equation}
which are equivalently expressed as
\begin{equation}
|H|\Lref\le h_f,
\qquad
|K|\Lref^2\le k_f .
\end{equation}
Thus, $h_f$ and $k_f$ specify what is treated as numerically indistinguishable
from zero after normalization by the hull reference length.

Using these thresholds, the valid curvature domain is partitioned as
\begin{align}
\mathcal{C}_{\mathrm{flat}}
&:
|K|\le\tau_K,\quad |H|\le\tau_H,
\\
\mathcal{C}_{\mathrm{single}}
&:
|K|\le\tau_K,\quad |H|>\tau_H,
\\
\mathcal{C}_{\mathrm{elliptic}}
&:
K>\tau_K,
\\
\mathcal{C}_{\mathrm{saddle}}
&:
K<-\tau_K .
\label{eq:curvature_classes}
\end{align}

The resulting categorical field $c(\mathbf{x})$ assigns the local curvature class to every valid point.
Let
$A_{\mathrm{flat}}$, $A_{\mathrm{single}}$,
$A_{\mathrm{elliptic}}$, and $A_{\mathrm{saddle}}$
denote the corresponding surface areas, and let $A_C$ denote the total valid
classified area. The global curvature-composition descriptor is then the vector
of area fractions
\begin{equation}
\mathbf{a}_C
=
\frac{1}{A_C}
\begin{bmatrix}
A_{\mathrm{flat}} &
A_{\mathrm{single}} &
A_{\mathrm{elliptic}} &
A_{\mathrm{saddle}}
\end{bmatrix}^\mathsf{T}.
\label{eq:classes}
\end{equation}
Its components sum to unity over the valid classified domain.

The class fractions describe the \emph{areal extent} of the different
curvature types rather than their intensity. For example, two hulls may have
similar global developability deviation while one concentrates strong double
curvature in a relatively small region and the other distributes weaker double
curvature over a much larger portion of the surface. The combination of
$I_D$, its signed components, and $\mathbf{a}_C$ therefore distinguishes curvature
intensity, sign, and spatial extent.

For the present assessment, fixed dimensionless values
\begin{equation}
h_f=k_f=10^{-4}
\end{equation}
are used consistently for all cases. 
Under the same Gaussian-curvature threshold and valid domain, the summed
elliptic and saddle fractions,
$a_{\mathrm{elliptic}}+a_{\mathrm{saddle}}$, give the total double-curved
area fraction. This threshold-dependent quantity is used below for the
robustness assessment but is not an additional component of the recommended
signature.

The classification retains the geometric connection to previous
curvature-based hull and plate classifications
\citep{Parsons1999,Kim2016}, but the adopted tolerances are numerical
quasi-zero definitions rather than manufacturing criteria. In particular,
they are not calibrated forming limits, fabrication-cost classes, or
process-selection boundaries.

\subsection{Geometry-based producibility signature}

The descriptors in the signature are designed to measure complementary rather than redundant aspects of hull geometry. Total and signed developability capture double-curvature intensity and sign, while curvature-class fractions describe its areal composition. These are genuinely different questions about the same surface, and collapsing them into a single weighted score would require process- or yard-specific calibration that is not available at the early design stage.

The recommended output is therefore a geometry-based producibility signature. For each retained metric family $k\in\mathcal K_{\mathrm{sig}}$, the signature contains
\begin{equation}
\mathcal P(\Ssurf)
=
\left\{
I_k,\,
q_k,\,
\nu_k
\right\}_{k\in\mathcal K_{\mathrm{sig}}},
\label{eq:signature}
\end{equation}
where $I_k$ is a global descriptor, $q_k$ is the corresponding distributed surface field, and $\nu_k$ records metric-specific validity and representation provenance. The global descriptor provides a compact quantity for comparing candidate geometries, the distributed output identifies where the response originates, and the validity record establishes whether the reported quantity is finite and numerically resolved.

The compact vector of retained global outcomes is
\begin{equation}
\mathbf{I}_{\mathrm{sig}}
=
\left[
I_D,\,
I_D^+,\,
I_D^-,\,
\mathbf{a}_C^\mathsf{T}
\right]^\mathsf{T}.
\label{eq:vector}
\end{equation}
The seven scalar entries in $\mathbf{I}_{\mathrm{sig}}$ span only five
independent degrees of freedom: $I_D = I_D^+ + I_D^-$ and the four class
fractions necessarily sum to unity. The two dependent entries are retained
intentionally---having $I_D$, $I_D^+$, and $I_D^-$ together makes the
intensity--sign breakdown immediately readable, while listing all four class
fractions preserves a complete and directly interpretable description of the
surface composition. Analysts using the signature in surrogate models,
dimensionality-reduction methods, or statistical workflows should account for
these built-in dependencies rather than treating all seven values as
independent inputs.
\begin{figure*}[!t]
\centering
\includegraphics[width=0.96\textwidth]{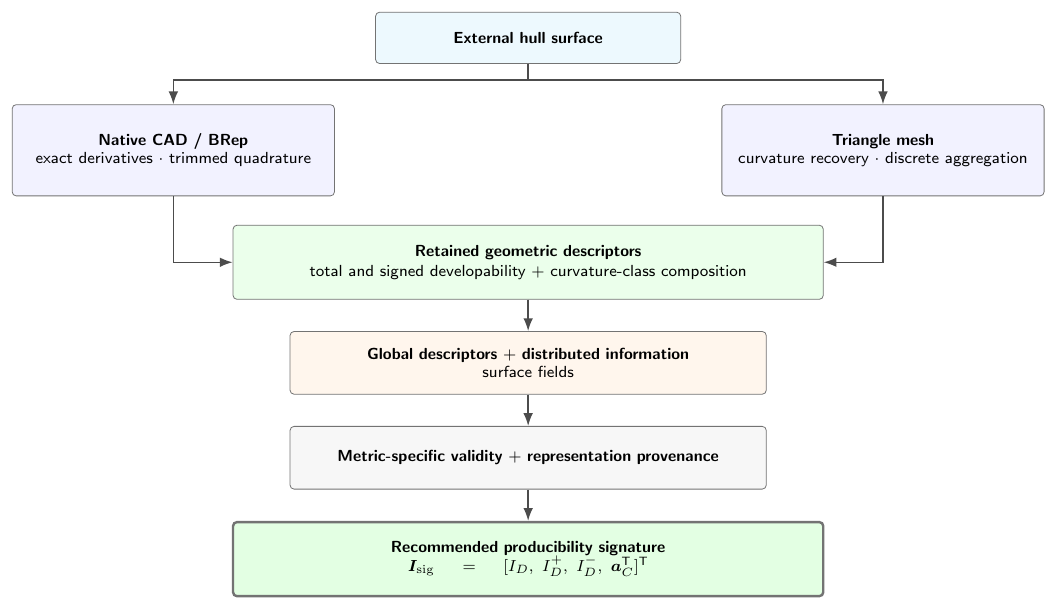}
\caption{Representation-aware workflow for the recommended geometry-based producibility signature. Native CAD/BRep and triangle-mesh paths evaluate the same retained descriptor families through representation-appropriate numerical operations. Global values and distributed surface fields are reported together with metric-specific validity and representation provenance; no universal composite score is introduced.}
\label{fig:signatureworkflow}
\end{figure*}

No composite producibility index or universal weighting is proposed. After appropriate physical or process-specific calibration, individual signature components or distributed fields may subsequently be used as objectives, constraints, surrogate responses, or features in design-space analysis, but those downstream applications are outside the present study.

\section{Representation-aware numerical realization}
\label{sec:numerics}

The continuous descriptors introduced in the previous section do not prescribe
a unique numerical implementation. Their realization depends on how the hull
geometry is represented. In particular, a native CAD surface provides access
to parametric derivatives of the underlying continuous geometry, whereas a
triangle mesh provides only discrete vertex positions, connectivity, and face
orientation from which differential quantities must be reconstructed.

The two representations are therefore treated as distinct numerical
realizations of the same conceptual descriptors. Native BRep evaluation is
used when IGES or STEP geometry is available; triangulated STL, OBJ, or PLY
geometry is evaluated through a discrete differential-geometry backend.
Neither representation is silently converted into the other for the purpose
of computing the retained metrics. In particular, a tessellation generated
from a CAD surface for visualization does not enter the native-BRep
integrals.

The combined workflow in Figure~\ref{fig:signatureworkflow} summarizes this
separation. Both numerical paths return global descriptors, distributed fields
and metric-specific validity and provenance
information. The purpose of this section is to describe the numerical
realization in sufficient detail for the metrics to be reproduced
independently of the accompanying software.

\subsection{Native BRep backend}
\label{sec:brep_backend}

Consider a parametric BRep face
\begin{equation}
\mathbf{S}(u,v)
:
\Omega \subset \mathbb{R}^2
\rightarrow
\mathbb{R}^3 ,
\label{eq:brep_surface}
\end{equation}
where $\Omega$ denotes the trimmed parameter domain of the face.
At a regular point, the first derivatives
$\mathbf{S}_u$ and $\mathbf{S}_v$ span the local tangent plane.
The coefficients of the first fundamental form are
\begin{equation}
E=\mathbf{S}_u\cdot\mathbf{S}_u,
\qquad
F=\mathbf{S}_u\cdot\mathbf{S}_v,
\qquad
G=\mathbf{S}_v\cdot\mathbf{S}_v,
\end{equation}
so that
\begin{equation}
\mathbf{I}
=
\begin{bmatrix}
E & F\\
F & G
\end{bmatrix},
\qquad
D
=
\det(\mathbf{I})
=
EG-F^2 .
\label{eq:brep_first_form}
\end{equation}

The corresponding surface-area Jacobian is
\begin{equation}
J(u,v)
=
\left\|
\mathbf{S}_u\times\mathbf{S}_v
\right\|
=
\sqrt{D},
\label{eq:brep_jacobian}
\end{equation}
and the oriented unit normal is
\begin{equation}
\mathbf{n}
=
\frac{
\mathbf{S}_u\times\mathbf{S}_v
}{
\left\|
\mathbf{S}_u\times\mathbf{S}_v
\right\|
},
\label{eq:brep_normal}
\end{equation}
with its orientation adjusted consistently with the topology of the
BRep face.

Using the second parametric derivatives, the coefficients of the
second fundamental form are
\begin{equation}
e
=
\mathbf{n}\cdot\mathbf{S}_{uu},
\qquad
f
=
\mathbf{n}\cdot\mathbf{S}_{uv},
\qquad
g
=
\mathbf{n}\cdot\mathbf{S}_{vv},
\end{equation}
or, equivalently,
\begin{equation}
\mathbf{II}
=
\begin{bmatrix}
e & f\\
f & g
\end{bmatrix}.
\label{eq:brep_second_form}
\end{equation}

The local shape operator is obtained from the first and second
fundamental forms as
\begin{equation}
\mathbf{B}
=
\mathbf{I}^{-1}\mathbf{II},
\label{eq:brep_shape_operator}
\end{equation}
whose eigenvalues are the principal curvatures
$\kappa_1$ and $\kappa_2$
\citep{doCarmo1976}.
Consequently,
\begin{equation}
H
=
\frac{1}{2}\operatorname{tr}(\mathbf{B})
=
\frac{eG-2fF+gE}{2D},
\end{equation}
\begin{equation}
K
=
\det(\mathbf{B})
=
\frac{eg-f^2}{D}.
\label{eq:brephk}
\end{equation}
Thus, the native BRep realization evaluates the same $H$ and $K$
introduced in Section~\ref{sec:framework} directly from the
differential geometry of the represented parametric surface.
The sign of $H$ depends on the selected surface orientation, whereas
$K$ is orientation invariant.

A point is not considered differentially valid when
$J$ or, equivalently, $D$ becomes degenerate.
Such a local failure does not by itself imply a singular physical
surface. A parameterization degeneracy, a measure-zero geometric
singularity, and a genuinely non-integrable curvature field are
distinguished through the metric-specific validity assessment
described in Section~\ref{sec:metric_validity}.

Once a local surface quantity $\phi(\mathbf{x})$ has been evaluated,
its integral is computed over the physical surface through the
parametric representation. For a BRep composed of faces
$\{\mathbf{S}_f\}$,
\begin{equation}
\int_{\mathcal{S}}
\phi(\mathbf{x})\,\mathrm{d}A
=
\sum_f
\int_{\Omega_f}
\phi\!\left(
\mathbf{S}_f(u,v)
\right)
J_f(u,v)\,
\mathrm{d}u\,\mathrm{d}v ,
\label{eq:brep_surface_integral}
\end{equation}
where $\Omega_f$ is the trimmed parameter domain of face $f$.
The trim topology is therefore part of the integration problem;
native-BRep integrals are not evaluated on a tessellated
approximation of the surface.

Numerically, the parameter domains are partitioned according to the
continuity intervals required by the evaluated differential quantity
and integrated using tensor-product Gauss--Legendre rules.
Adaptive subdivision is used to concentrate quadrature effort in
parameter-space regions requiring additional resolution
\citep{DavisRabinowitz1984,Cools2003}.
For each integration cell, lower- and higher-order quadrature
estimates are compared. A cell is subdivided when the relative
discrepancy exceeds the prescribed tolerance or when additional
resolution of the trimmed domain is required. Refinement continues
until the convergence criterion is satisfied or the declared maximum
depth is reached.

Failure to satisfy the requested tolerance is retained explicitly as
numerical-convergence evidence rather than hidden by accepting the
last finite quadrature value. This distinction is required because a
finite-depth result may correspond to a convergent integral, an
unresolved integral, or a discretization-dependent cutoff near a
geometric singularity.

As independent numerical checks, the represented surface area is also
evaluated using the CAD kernel's geometric-property operator.
Moreover, the values of $H$ and $K$ obtained from
Eq.~\eqref{eq:brephk} are cross-checked pointwise against an
independent CAD differential-property evaluator. These checks are
used for verification of the implementation and do not replace the
geometric formulation above.

\subsection{Triangle-mesh backend}
\label{sec:mesh_backend}

Let the triangulated representation of the hull surface be
\begin{equation}
\mathcal{S}_h=(\mathcal V,\mathcal F),
\end{equation}
where
\begin{equation}
\mathcal V
=
\{\mathbf{x}_i\}_{i=1}^{N_v}
\end{equation}
is the set of $N_v$ vertex positions and
$\mathcal F$ is the set of oriented triangular faces.
Each face $f=(i,j,k)\in\mathcal F$ is therefore defined by three
vertex indices and the corresponding positions
$\mathbf{x}_i$, $\mathbf{x}_j$, and $\mathbf{x}_k$.

Unlike a native BRep, a triangle mesh does not provide parametric
first and second derivatives of the represented surface. Differential
quantities must instead be reconstructed from the local variation of
the discrete geometry. The present realization follows the
curvature-tensor method of \citet{Rusinkiewicz2004}.

For a sufficiently smooth surface, the shape operator maps tangent
directions to the corresponding variation of the surface normal and,
in an orthonormal tangent basis, is represented by a symmetric
$2\times2$ tensor. Its eigenvalues are the principal curvatures.
The mesh realization approximates this same local object from
discrete normal variations.

Consistently oriented vertex normals $\mathbf{n}_i$ are first constructed
from the geometry of the incident triangles. For each triangular face,
a local orthonormal tangent basis is introduced. The changes of the
vertex normals along the independent edge directions are then fitted
to a symmetric curvature tensor,
\begin{equation}
\widehat{\mathbf{B}}_f
=
\begin{bmatrix}
b_{11} & b_{12}\\
b_{12} & b_{22}
\end{bmatrix},
\end{equation}
which provides a discrete approximation of the local shape operator
on face $f$.

The face tensors incident to a vertex are subsequently transported
into a common tangent frame at that vertex and accumulated using
geometric corner-area weights, following the procedure of
\citet{Rusinkiewicz2004}. The resulting vertex tensor is denoted by
$\widehat{\mathbf{B}}_i$. Its eigendecomposition,
\begin{equation}
\widehat{\mathbf{B}}_i\,\mathbf{t}_{\alpha,i}
=
\kappa_{\alpha,i}\,\mathbf{t}_{\alpha,i},
\qquad
\alpha=1,2,
\end{equation}
provides the reconstructed principal curvatures
$\kappa_{1,i}$ and $\kappa_{2,i}$ and their corresponding principal
directions $\mathbf{t}_{1,i}$ and $\mathbf{t}_{2,i}$.

The discrete mean and Gaussian curvatures are therefore
\begin{equation}
H_i
=
\frac{\kappa_{1,i}+\kappa_{2,i}}{2},
\qquad
K_i
=
\kappa_{1,i}\kappa_{2,i}.
\label{eq:mesh_hk}
\end{equation}
Reversal of the mesh orientation reverses the signs of the principal
curvatures and hence of $H_i$, whereas $K_i$ remains unchanged.
This is consistent with the continuous definitions used for the
native-BRep realization.

To approximate surface integrals, each vertex is assigned an
associated surface area. Let
\begin{equation}
\mathcal F_i
=
\{f\in\mathcal F : i\in f\}
\end{equation}
denote the set of faces incident to vertex $i$, and let $A_f$ be the
area of face $f$. The barycentric area associated with vertex $i$ is
\begin{equation}
A_i
=
\frac{1}{3}
\sum_{f\in\mathcal F_i}
A_f .
\label{eq:vertex_area}
\end{equation}
Thus, each triangle contributes one third of its area to each of its
three vertices, and
\begin{equation}
\sum_{i\in\mathcal V}A_i
=
\sum_{f\in\mathcal F}A_f .
\end{equation}

At every vertex where Gaussian curvature is considered valid, the
discrete counterpart of the local developability-deviation field is
\begin{equation}
q_{D,i}
=
|K_i|L_{\mathrm{ref}}^2 .
\end{equation}
Let $\mathcal V_K\subseteq\mathcal V$ denote the subset of vertices
for which the reconstructed Gaussian curvature is valid. The mesh
realization of the corresponding global descriptor is then
\begin{equation}
I_D^h
=
\frac{
\displaystyle
\sum_{i\in\mathcal V_K}
A_i\,q_{D,i}
}{
\displaystyle
\sum_{i\in\mathcal V_K}
A_i
}.
\label{eq:mesh_ID}
\end{equation}
The signed developability components and curvature-class area
fractions are aggregated analogously from their corresponding local
fields over their declared valid domains.

Boundary vertices are excluded from the retained Gaussian-curvature
domain. At an open mesh boundary, the local neighborhood is
incomplete; applying the same interior reconstruction without an
explicit boundary model would mix a representation-specific edge
treatment with the intended surface quantity. Boundary samples are
therefore marked invalid for retained $K$-based aggregation rather
than being assigned zero curvature. The corresponding represented
and valid areas are reported explicitly.

Changes of adjacent triangle normals are additionally retained as a
mesh-quality and representation diagnostic. Such quantities can
identify coarse, irregular, or locally defective tessellation
regions, but they depend directly on mesh resolution, connectivity,
and element quality and have no equivalent role in the native-BRep
signature. They are therefore not included in the recommended
producibility signature.

No smoothing, percentile clipping, hidden remeshing, or automatic
geometry repair is applied to force numerical agreement with the
native-BRep realization. Mesh topology, boundary fraction, triangle
quality, edge-length distribution, associated-area distribution,
vertex valence, connected components, and watertightness are instead
retained as representation provenance and numerical-quality
information.

\subsection{Metric-specific validity}\label{sec:metric_validity}

A successful numerical evaluation does not imply that every retained quantity
is mathematically meaningful or numerically resolved on the represented
geometry. Surface area requires a regular area measure, Gaussian-curvature
descriptors require the corresponding differential information and integrable
surface fields.
Validity is therefore attached to each metric rather than to the hull geometry
as a whole. For every reported quantity, the numerical realization retains the
metric value when finite, the represented and valid areas, the backend and
representation, and the available convergence or regularity evidence. A metric
that cannot be established as finite is reported with an explicit null value
rather than with the last finite number returned by a discretization or
quadrature rule.

Three situations are particularly important. First, an isolated
parameterization degeneracy does not necessarily imply a singular physical
surface; the metric may remain finite when expressed in physical coordinates.
Second, a measure-zero geometric singular set may still admit a convergent
improper integral, in which case the corresponding global metric remains
well-defined but its singular provenance is retained. Third, the integral
itself may diverge as the singular neighborhood is resolved. In that case a
finite value obtained at a particular quadrature depth or mesh resolution is a
numerical cutoff and is not promoted to a reported metric.

Similarly, lack of numerical convergence is distinguished from mathematical
divergence. If the available refinement is insufficient to establish either
convergence or divergence, the quantity remains explicitly unresolved. For
mesh realizations, material dependence on resolution, connectivity, or mesh
quality is retained as representation-sensitivity evidence rather than removed
by smoothing or clipping.

The resulting validity vocabulary distinguishes finite and converged values,
convergent improper integrals, cautionary measure-zero singularities,
unresolved quadrature, parameterization degeneracy, non-integrable geometric
singularities, insufficient surface continuity, representation-sensitive mesh
results, quantities not evaluated, and quantities not applicable to a given
representation. The detailed numerical tolerances and benchmark-specific
settings used in the present assessment are reported separately in
Section~\ref{sec:verification}.

Validity is assigned per metric, not per geometry. Public display labels
summarize, but do not replace, the detailed scientific classification:
\texttt{VALID} denotes an established usable value, and \texttt{VALID*} a
convergent improper integral whose singular provenance remains explicit;
\texttt{CAUTION} denotes a finite result requiring a stated qualification,
such as a measure-zero singular set or parameterization degeneracy;
\texttt{UNCONVERGED} means that bounded refinement did not establish
convergence, so a finite-depth candidate is not an established metric value;
and \texttt{SINGULAR} denotes a non-integrable geometric singularity.
\texttt{MESH-SENSITIVE} identifies a mesh result dependent on resolution,
connectivity, or quality. Detailed reasons, codes, valid area, convergence
evidence, and provenance remain part of the record.

\section{Verification and benchmark protocol}
\label{sec:verification}

The verification protocol first uses analytical and semi-analytical controls to verify the continuous metric definitions and their numerical realization on geometries with known curvature properties. Three numerical questions are then kept distinct: controlled same-face BRep-to-mesh comparisons isolate curvature-recovery and connectivity effects on matched geometry; the frozen coarse/medium/fine STL campaign measures whole-hull mesh-resolution sensitivity; and native-BRep versus frozen fine-STL comparisons assess representation sensitivity. The complete framework is exercised on the public hull geometries together with metric-specific regularity and validity evidence.

\subsection{Analytical and semi-analytical controls}

\begin{figure*}[!b]
\centering
\includegraphics[width=\textwidth]{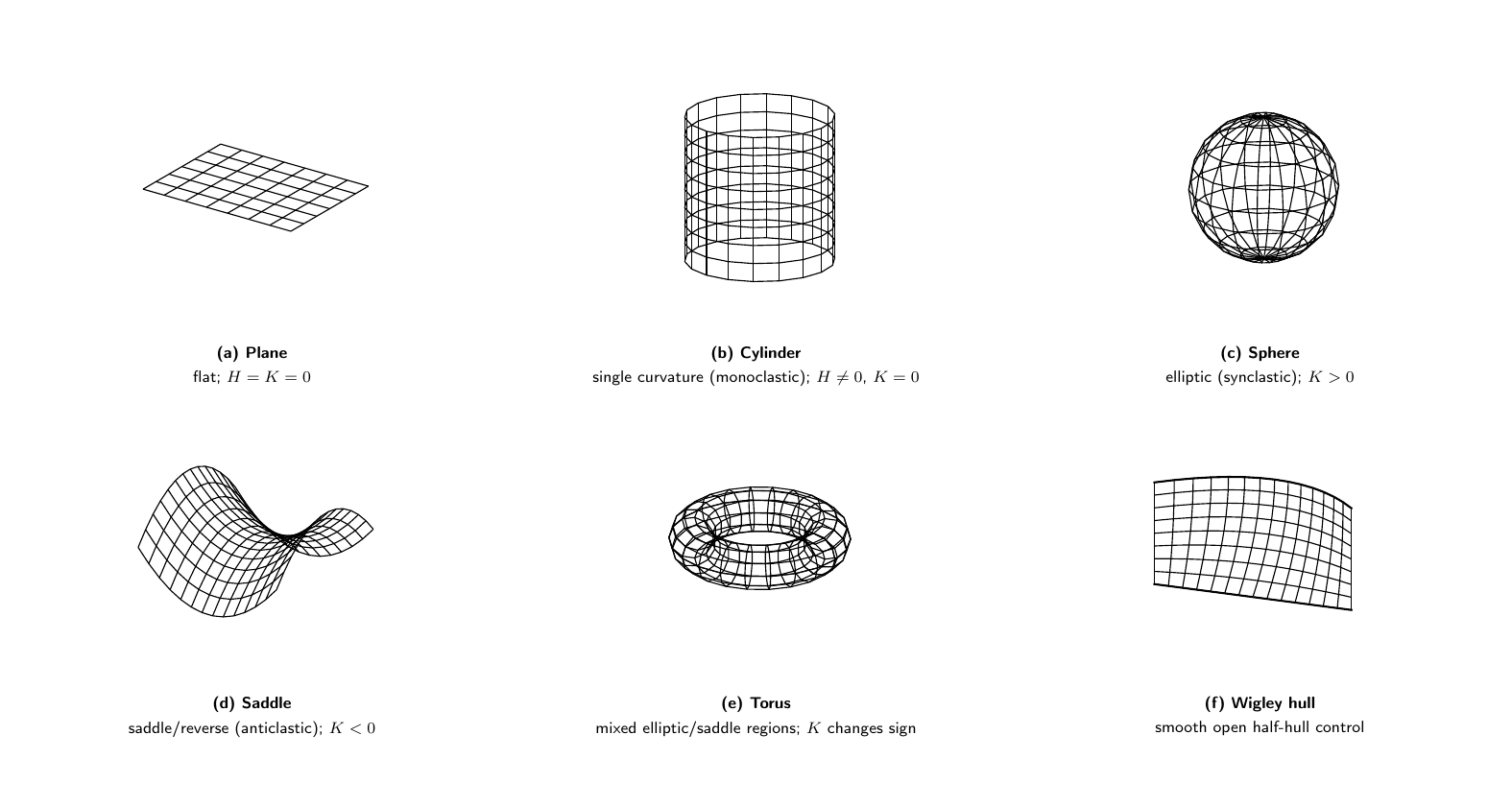}
\caption{Analytical and semi-analytical controls used before the public-hull campaign.}
\label{fig:controls}
\end{figure*}

The analytical sequence comprises a plane, sphere, cylinder, saddle, torus, and half-Wigley hull, shown in Fig.~\ref{fig:controls}. These geometries progressively exercise different features of the formulation. The plane provides the zero-curvature and boundary control; the sphere has constant nonzero $H$ and positive $K$; the cylinder isolates single curvature with $K=0$; and the saddle provides a negative-$K$ control. The torus introduces smooth spatially varying curvature on a closed surface, whereas the half-Wigley hull provides a smooth open, hull-like control with an analytical polynomial surface definition, enabling independent semi-analytical evaluation of the retained curvature-based descriptors.

The plane, sphere, cylinder, and torus admit exact reference quantities, whereas the half-Wigley hull admits an analytical differential-geometry description from which independent semi-analytical reference values can be obtained. The half-Wigley surface is defined as
\begin{equation}
y=\frac{B}{2}
\left[1-\left(\frac{2x}{L}\right)^2\right]
\left[1-\left(\frac{z}{T}\right)^2\right],
\end{equation}
for $-L/2\le x\le L/2$ and $-T\le z\le0$, where $L$, $B$, and $T$ denote length, breadth, and draft, respectively. Here, $L=1$, $B=0.1$, $T=0.0625$ in consistent length units, and $\Lref=L$. Analytical surface derivatives provide $H$, $K$, and the area Jacobian; the resulting surface integrals are evaluated using independent high-accuracy numerical quadrature to obtain reference values for
$I_D$, $I_D^\pm$, and the threshold-based curvature-class fractions.
These references are reported together with the corresponding native-BRep results in Table~\ref{tab:brepanalytic}.
The torus, with major radius $R=3$, minor radius $r=1$, and $\Lref=8$, additionally provides a nontrivial reference for surface-integrated Gaussian curvature. Native quadrature is assessed through integration order and bounded refinement. The mesh realization is evaluated independently on consistent, alternating, and deterministic random diagonal patterns of the same underlying geometries, allowing sensitivity to connectivity to be distinguished from changes in shape.

\subsection{Public-hull campaign and representation verification}
\begin{figure*}[!t]
\centering
\includegraphics[width=\textwidth]{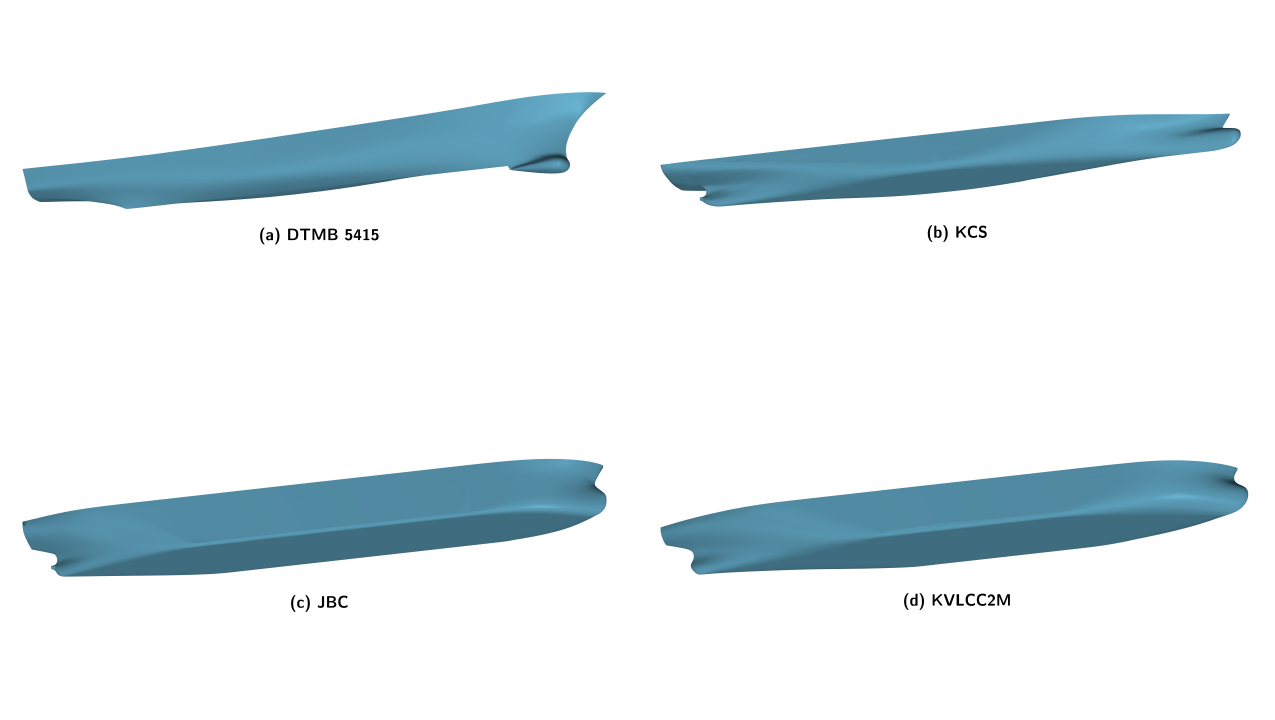}
\caption{Representative public-hull geometries used in the assessment: (a) DTMB 5415, (b) KCS, (c) JBC, and (d) the NMRI faired KVLCC2M. The four panels preserve hull proportions and use one common orthographic viewing convention.}
\label{fig:publicsuite}
\end{figure*}

The public-hull campaign is restricted to four distinct benchmark
geometries: DTMB 5415, KCS, JBC, and KVLCC2M. The cases span naval and
commercial hull forms and different native source representations, and are
used to assess the behavior, complementarity, representation sensitivity, and
metric-specific validity of the proposed descriptors rather than to rank ship
types. Figure~\ref{fig:publicsuite} introduces the four geometries using a
common viewing convention, while Table~\ref{tab:publicprovenance} summarizes
the adopted reference-length conventions, model scales, and authoritative
source representations.

Controlled exact-same-face pointwise BRep-to-mesh curvature verification is
performed for all four public-hull cases. For each case, an unchanged
authoritative BRep face is sampled on a structured parameter grid, and the
comparison-mesh vertices are placed directly at the corresponding
$\mathbf{S}(u,v)$ locations; no nearest-neighbor projection is used. Grid
levels of $12\times12$, $24\times24$, and $48\times48$ panels are evaluated
over the same inset parameter region. Consistent and alternating
triangle-diagonal patterns are considered, and a common two-ring
strict-interior filter excludes incomplete boundary neighborhoods. This
construction isolates curvature-recovery and connectivity effects from
CAD-to-mesh correspondence errors.

For $X\in\{H,K\}$, let $\mathbf{x}^{h}$ and
$\mathbf{x}^{\mathrm{BRep}}$ collect the recovered and direct-BRep values at
the identical physical vertices, and let
$\boldsymbol{\delta}_X=\mathbf{x}^{h}-\mathbf{x}^{\mathrm{BRep}}$. After
zeroing both compared vectors outside the common retained mask,
curvature-recovery accuracy is measured using
\begin{equation}
\varepsilon_X
=
\left[
\frac{\boldsymbol{\delta}_X^\mathsf{T}\mathbf{M}\boldsymbol{\delta}_X}
{(\mathbf{x}^{\mathrm{BRep}})^\mathsf{T}\mathbf{M}
\mathbf{x}^{\mathrm{BRep}}}
\right]^{1/2},
\qquad X\in\{H,K\},
\label{eq:sameface_error}
\end{equation}
where $\mathbf{M}$ is the standard consistent P1 surface mass matrix assembled
over the comparison triangulation using the physical triangle areas.
The common retained mask requires the two-ring strict interior together with
valid, finite recovered and reference values of both $H$ and $K$. A single
global sign for recovered $H$ is selected by correlation with the signed BRep
reference; $K$ requires no sign alignment. The global-norm denominator avoids
unstable pointwise relative division near zero. Absolute median and
95th-percentile errors are retained in the reproducibility data but are not
used for the main comparison.

For the four cases, the reported native results use
$L_{\mathrm{ref}}=L_{pp}$ and the common numerical quasi-zero factors
$h_f=k_f=10^{-4}$. Here $L_{pp}$ is a global hull-scale normalization for
dimensionless early-stage comparison, not the characteristic length of an
individual plate or forming operation. Native trimmed-domain integration uses tensor-product
Gauss--Legendre order 5, relative tolerance $10^{-4}$, and nominal maximum
refinement depth 5. When convergence is not established at this depth, the
check is repeated with maximum depth 6; quantities that remain unresolved are
reported as such rather than being assigned the final finite-depth value.
The $10^{-4}$ tolerance governs local adaptive-cell quadrature refinement,
whereas public metric validity follows bounded global-refinement evidence with
metric-specific stability criteria; for retained $I_D$, the final and prior
relative changes must not exceed 2\% and 3\%, respectively.

Whole-hull mesh-resolution sensitivity is additionally assessed
using the frozen coarse, medium, and fine campaign tessellations. Changes in
the curvature-class composition between two mesh levels $r_1$ and $r_2$ are
summarized by
\begin{equation}
d_1\!\left(\mathbf{a}_C^{(r_1)},\mathbf{a}_C^{(r_2)}\right)
=
\left\|\mathbf{a}_C^{(r_1)}-\mathbf{a}_C^{(r_2)}\right\|_1,
\label{eq:class_l1}
\end{equation}
i.e., the sum of the absolute changes of the four class fractions.

All geometries are evaluated at their documented model scale. The half- or
full-hull distinction reported in Table~\ref{tab:publicprovenance} records the
authoritative source representation and is retained as provenance; it is not
used as a basis for cross-hull comparison. The descriptors apply to the
represented assessed surface. Additional surfaces such as decks or appendages,
when included in the supplied geometry, alter the area normalization and
curvature composition and are therefore part of the representation provenance.
Detailed source hashes, coordinate
transforms, source scales, source-unit information, numerical settings, and
comparison statistics are retained in the machine-readable provenance
accompanying the benchmark results.

\begin{table*}[!t]
\centering
\caption{Reference-length conventions, model scales, and authoritative
source representations for the four public-hull cases used in the
assessment.}
\label{tab:publicprovenance}
\begin{tabular}{llll}
\toprule
Case & $L_{\mathrm{ref}}$ [m] & Model scale & Authoritative source representation \\
\midrule
DTMB 5415 & 5.72   & $1{:}24.8240$  & starboard half hull \\
KCS       & 7.2786 & $1{:}31.5994$ & starboard half bare hull \\
JBC       & 7.00   & $1{:}40.0000$      & full symmetric bare hull \\
KVLCC2M   & 4.97   & $1{:}64.3900$  & NMRI faired full symmetric bare hull \\
\bottomrule
\end{tabular}
\end{table*}

\section{Results}
\label{sec:results}

The results assess the proposed geometry-based producibility framework
from three complementary perspectives. First, analytical and semi-analytical controls are
used to verify the continuous definitions and their numerical
realizations. Second, the principal descriptors are evaluated on the
public-hull set to examine their global behavior and distributed
geometric interpretation. Finally, the numerical evidence separately
addresses matched-geometry curvature recovery, whole-hull STL
mesh-resolution sensitivity, and native-BRep/fine-STL representation
sensitivity.

The assessment is intended as a proof of concept of the proposed
geometry-based signature rather than as a ranking of hull
producibility. Accordingly, the results are interpreted in terms of
complementarity, spatial localization, and numerical robustness of the
individual descriptors. Developability, curvature-class composition,
and their distributed surface fields form the public-hull signature; their numerical
realization is interpreted together with metric-specific mathematical
and numerical validity.

\subsection{Analytical and semi-analytical verification}
\label{sec:results_verification}

Table~\ref{tab:brepanalytic} summarizes direct BRep verification of the
recommended signature. The plane and cylinder recover zero total
and signed developability together with the expected flat and
single-curvature classes. The sphere recovers positive developability
and an entirely elliptic surface, while the saddle recovers negative
developability and an entirely saddle-class valid domain. The torus
provides a mixed-sign control and verifies trimmed-domain quadrature
independently of a triangle curvature field.

\begin{table*}[!t]
\centering
\caption{Native-BRep verification of the recommended geometry-based producibility signature. Class vectors are ordered as (flat, single, elliptic, saddle). Errors are relative for nonzero scalar references; for the paired signed quantities, the maximum relative component error is reported, whereas for curvature-class vectors the maximum absolute component error is reported.}
\label{tab:brepanalytic}
\small
\begin{tabular}{lllll}
\toprule
Control & Quantity & Reference & Native BRep result & Error \\
\midrule

\multirow{3}{*}{Plane}
& $I_D$
& $0$
& $0$
& $0$ \\

& $I_D^+/I_D^-$
& $0/0$
& $0/0$
& $0$ \\

& $\mathbf{a}_C$
& $(1,0,0,0)$
& $(1,0,0,0)$
& $0$ \\
\cmidrule(lr){2-5}

\multirow{3}{*}{Unit sphere}
& $I_D$
& $4.00000000$
& $4.00000000$
& $2.22\times10^{-16}$ \\

& $I_D^+/I_D^-$
& $4/0$
& $4/0$
& $2.22\times10^{-16}$ \\

& $\mathbf{a}_C$
& $(0,0,1,0)$
& $(0,0,1,0)$
& $0$ \\
\cmidrule(lr){2-5}

\multirow{3}{*}{Cylinder}
& $I_D$
& $0$
& $0$
& $0$ \\

& $I_D^+/I_D^-$
& $0/0$
& $0/0$
& $0$ \\

& $\mathbf{a}_C$
& $(0,1,0,0)$
& $(0,1,0,0)$
& $0$ \\
\cmidrule(lr){2-5}

\multirow{3}{*}{Saddle}
& $I_D$
& $I_D>0$ (sign)
& $0.746436$
& -- \\

& $I_D^+/I_D^-$
& $0/I_D$ (sign)
& $0/0.746436$
& -- \\

& $\mathbf{a}_C$
& $(0,0,0,1)$
& $(0,0,0,1)$
& $0$ \\
\cmidrule(lr){2-5}

\multirow{3}{*}{Torus}
& $I_D$
& $13.58122181$
& $13.58122181$
& $2.62\times10^{-16}$ \\

& $I_D^+/I_D^-$
& $6.79061091/6.79061091$
& $6.79061091/6.79061091$
& $3.92\times10^{-16}$ \\

& $\mathbf{a}_C$
& $(0,\,2.9842\times10^{-6},\,0.606102,\,0.393895)$
& $(0,\,0,\,0.606103,\,0.393897)$
& $2.98\times10^{-6}$ \\
\cmidrule(lr){2-5}

\multirow{3}{*}{Half-Wigley}
& $I_D$
& $4.30990339$
& $4.30989710$
& $1.46\times10^{-6}$ \\

& $I_D^+/I_D^-$
& $2.15495169/2.15495169$
& $2.15494850/2.15494850$
& $1.48\times10^{-6}$ \\

& $\mathbf{a}_C$
& $(0,\,5.9863\times10^{-5},\,0.626187,\,0.373753)$
& $(0,\,0,\,0.625883,\,0.374117)$
& $3.64\times10^{-4}$ \\

\bottomrule
\end{tabular}
\end{table*}

The torus provides exact nonzero references for both signed developability and threshold-based curvature classification. Its geometry yields equal positive and negative absolute-Gaussian-curvature contributions, giving $I_D^+=I_D^-=64/(3\pi)$. In addition, the region classified as quasi-zero Gaussian curvature under the adopted threshold occupies $2.9842\times10^{-6}$ of the total surface area. The native-BRep results recover both references within the prescribed quadrature tolerance.
The half-Wigley control provides an additional hull-like mixed-curvature verification case. Independent semi-analytical integration gives $I_D=4.30990339$, $I_D^+=2.15495169$, and $I_D^-=2.15495169$. The native-BRep realization recovers these quantities with relative errors of approximately $1.5\times10^{-6}$, while the threshold-based curvature-class composition is recovered with a maximum absolute component error of $3.64\times10^{-4}$.

Additional controls verify scale invariance, orientation reversal,
smooth face partitioning, trimmed circular caps, sharp dihedral features
with positional but not tangent continuity ($C^0$), open-surface handling,
and independence of native-BRep values from display tessellation.

\paragraph{Mesh realization.}
The selected Rusinkiewicz reconstruction agrees with the sphere, cylinder, and torus references within the verification tolerance under consistent, alternating, and deterministic random diagonal patterns. 
The half-Wigley control additionally exercises boundary handling on a smooth open hull-like surface.
These controls verify the selected mesh curvature realization while confirming that derivative-based quantities remain sensitive to numerical representation and connectivity.

\subsection{Public-hull global signatures}
\label{sec:results_global}

Table~\ref{tab:publicsignatures} reports the complete global signature for both the native-BRep and frozen fine-STL representations of each public hull. The native-BRep rows provide the reference global values for the adopted authoritative BRep used for the cross-hull geometric interpretation in this subsection; the fine-STL rows are reported alongside them to expose representation sensitivity, which is examined explicitly in Section~\ref{sec:results_representation}. The signature components are reported individually and are not combined into an uncalibrated universal producibility score. 
The frozen fine-STL representations were produced from the corresponding
campaign CAD sources through the documented OpenCascade conversion and
tessellation workflow, using a common linear deflection
$8.75\times10^{-4}\Lref$ and angular deflection $0.2$ rad. The resulting meshes contain 3,110, 9,643, 5,629, and 4,096 triangles for DTMB~5415, KCS, JBC, and KVLCC2M, respectively. They are retained unchanged throughout the representation- and mesh-resolution-sensitivity analyses.

\begin{table*}[!b]
\centering
\caption{Geometry-based producibility signatures for the four public hulls
from native BRep and frozen fine-STL representations. Relative differences
$\Delta I_D$ are computed for the fine-STL realization with respect to native
BRep. Curvature-class fractions are normalized on the valid area of the
corresponding representation. }
\label{tab:publicsignatures}
\setlength{\tabcolsep}{3.2pt}
\begin{tabular}{llrrrrrrr}
\toprule
Hull & Representation &
$I_D$ &
$\Delta I_D$ [\%] &
$I_D^+/I_D^-$ &
Flat &
Single &
Elliptic &
Saddle \\
\midrule

\multirow{2}{*}{DTMB 5415}
& Native BRep
& $^{*}$66.4011
& --
& 33.6190 / 32.7821
& 0.00000
& 0.00003
& 0.51225
& 0.48772 \\
& Fine STL
& 77.8628
& +17.3
& 40.8474 / 37.0154
& 0.00000
& 0.00000
& 0.48040
& 0.51960 \\
\cmidrule(lr){2-9}

\multirow{2}{*}{KCS}
& Native BRep
& 82.6084
& --
& 41.1121 / 41.4963
& 0.21266
& 0.05839
& 0.32758
& 0.40138 \\
& Fine STL
& 89.5729
& +8.4
& 51.1308 / 38.4421
& 0.00562
& 0.04471
& 0.35700
& 0.59267 \\
\cmidrule(lr){2-9}

\multirow{2}{*}{JBC}
& Native BRep
& $^{*}$58.6029
& --
& 22.6040 / 35.9989
& 0.50066
& 0.15527
& 0.15515
& 0.18892 \\
& Fine STL
& 72.8435
& +24.3
& 35.1914 / 37.6520
& 0.18783
& 0.01561
& 0.39963
& 0.39694 \\
\cmidrule(lr){2-9}

\multirow{2}{*}{KVLCC2M}
& Native BRep
& $^{*}$53.0006
& --
& 31.6765 / 21.3241
& 0.54009
& 0.02490
& 0.27006
& 0.16495 \\
& Fine STL
& 44.5400
& -16.0
& 27.7997 / 16.7403
& 0.23711
& 0.03766
& 0.34602
& 0.37922 \\

\bottomrule
\multicolumn{9}{l}{$^{*}$denotes a convergent improper
native-BRep developability integral.}
\end{tabular}
\end{table*}

Metric-specific validity qualifies how the corresponding signature values
should be interpreted numerically.

For each starred hull, a separate nested-neighborhood audit examines the
identified degenerate root set used for the improper-integral classification
and reports the removed physical area. The absolute-$K$
integral over the shrinking neighborhood decays to zero with fitted positive
powers of 0.97, 1.00, and 1.06 for DTMB 5415, JBC, and KVLCC2M,
respectively; the final relative changes in the complementary global $I_D$
sequence are 0.0512\%, 0.000670\%, and 0.0146\%. 
This refinement-based convergence evidence supports the
convergent-improper classification rather than relying only on agreement
between two terminal quadrature depths.

The results demonstrate complementarity rather than an ordering of
production difficulty. KCS has
approximately 24\% larger total developability deviation than DTMB 5415, while elliptic and saddle regions together
cover approximately 72.9\% of the KCS valid surface versus essentially
the entire DTMB valid surface. A hull may therefore exhibit stronger
average double-curvature intensity over a smaller portion of its
surface, while another exhibits lower average intensity distributed
over almost its entire valid area.

The signed components provide additional geometric information. DTMB
and KCS are nearly balanced between elliptic and saddle/reverse
developability intensity, KVLCC2M presents a larger positive than
negative contribution, and JBC presents the opposite behavior. The
curvature-class fractions complement these integral measures by
quantifying areal extent. Elliptic and saddle regions together cover
essentially the entire represented DTMB surface; the corresponding
double-curved fractions are approximately 72.9\% for KCS, 43.5\% for
KVLCC2M, and 34.4\% for JBC. Intensity, sign, and areal composition
therefore provide related but distinct information.

The total double-curved area fraction and its qualitative cross-hull ordering
remain stable across the threshold study
$h_f=k_f=10^{-5}$, $10^{-4}$, and $10^{-3}$.
Between the smallest and largest thresholds, the total double-curved area
fraction decreases by approximately $0.05\%$ for DTMB 5415, $2.2\%$ for KCS,
$5.0\%$ for JBC, and $0.84\%$ for KVLCC2M.
Individual flat and single-curvature fractions can be more
threshold-sensitive because the quasi-zero threshold directly controls their
classification. The values of $I_D$, $I_D^+$, and $I_D^-$ are independent of
these classification thresholds.

Additional higher-regularity curvature functionals based on area-averaged
$H^2$ and $|\gradS H|^2$, together with section-curvature variation and
spectral diagnostics, were investigated during metric development. The
higher-order surface functionals encountered regularity or integrability
limitations, while the section-based candidates showed CAD-join,
resampling, or spectral-scale sensitivity. Robust, representation-independent
general hull-level values were therefore not established for these candidate
families, and they were excluded from the recommended signature.

\subsection{Distributed localization of the main descriptors}
\label{sec:results_distributed}

\begin{figure*}[!t]
\centering
\includegraphics[width=\textwidth]{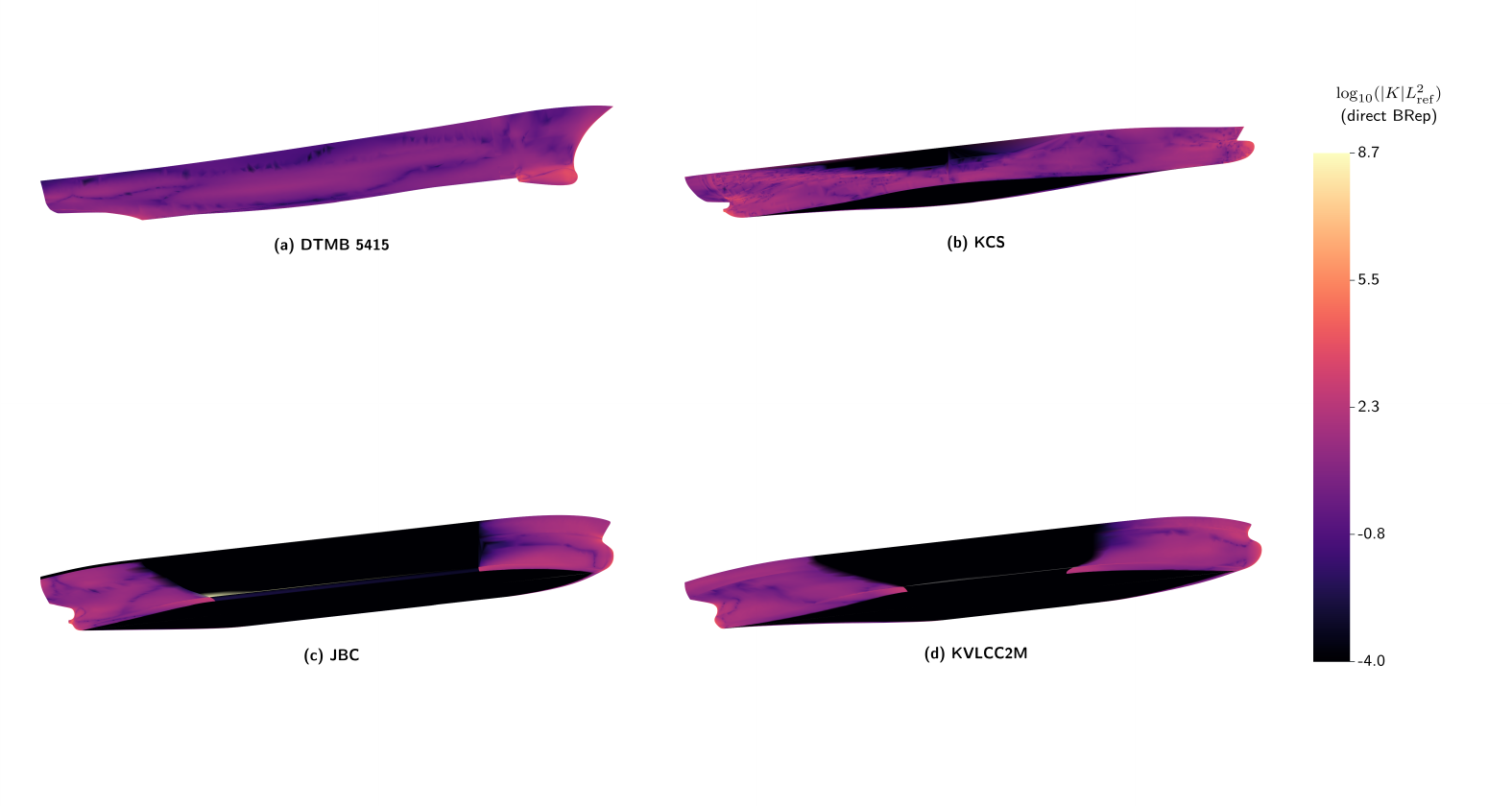}
\caption{Distributed native-BRep developability density $|K|\Lref^2$ for DTMB 5415, KCS, JBC, and KVLCC2M. Colors show direct BRep differential samples on display-only tessellations using the common robust display range $-4\le\log_{10}(|K|\Lref^2)\le8.692$; values outside this range are graphically saturated for visualization only, and invalid samples are grey. No clipping or saturation is applied to numerical integration or to the reported metrics. The lower display limit coincides with the declared quasi-zero scale $k_f=10^{-4}$, while the upper limit is the pooled 99.5th percentile of the positive samples. The corresponding global descriptor $I_D$ is the represented-valid-area average.}
\label{fig:developabilitydensity}
\end{figure*}

\begin{figure*}[!t]
\centering
\includegraphics[width=\textwidth]{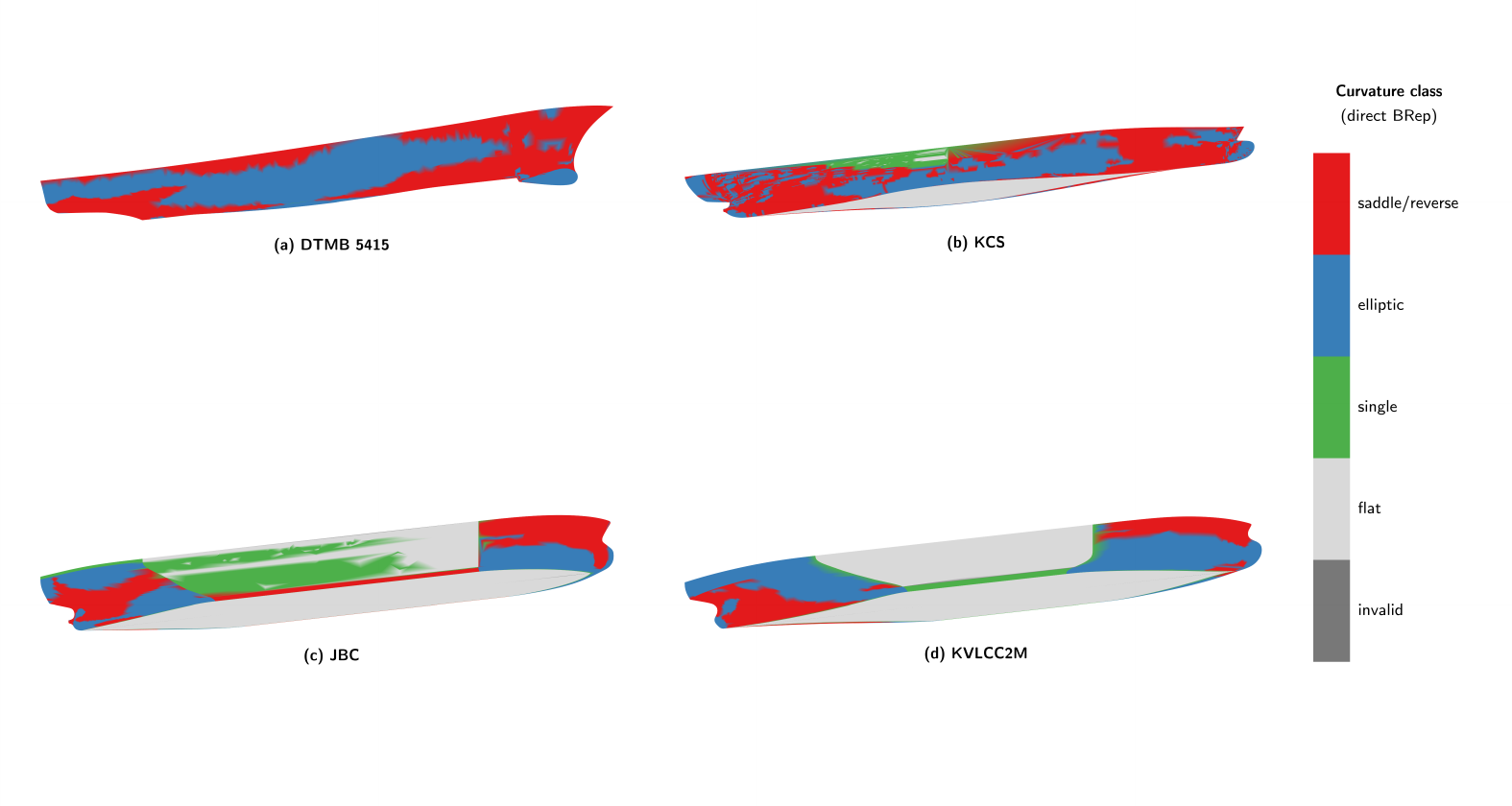}
\caption{Distributed native-BRep curvature classes for DTMB 5415, KCS, JBC, and KVLCC2M: invalid, flat, single curvature, elliptic double curvature, and saddle/reverse double curvature. The native-BRep fractions reported in Table~\ref{tab:publicsignatures} are the represented-valid-area aggregation of this categorical field. These are geometric classes under the declared quasi-zero thresholds, not calibrated forming-cost categories.}
\label{fig:classpanels}
\end{figure*}

The native-BRep global values in Table~\ref{tab:publicsignatures} do not identify
where the corresponding geometric contributions originate.
Figures~\ref{fig:developabilitydensity} and~\ref{fig:classpanels}
provide the distributed counterparts of the retained
signature for the same four public hulls. The surface panels draw direct
BRep differential samples on display-only tessellations; the
tessellations do not define the native-BRep integrals.

\paragraph{Developability intensity.}
For all cases, native developability is finite or a
convergent improper integral. $I_D$ is the represented-valid-area
average of $|K|\Lref^2$, and Fig.~\ref{fig:developabilitydensity}
localizes the regions responsible for the global values, including
contributions associated with the bow, stern, bilge, keel, and
individual CAD patches.

The positive-sample logarithmic field spans $-57.522$ to $35.288$. The extreme positive tail is
driven by isolated samples near degenerate or boundary-degenerate BRep
locations. At the smallest audited exclusions, the remaining contributions to
the full developability numerator are 0.05193\%, 0.0006695\%, and 0.01271\%
for DTMB 5415, the audited JBC set, and KVLCC2M, respectively, and decay under
neighborhood refinement; these audited local extremes therefore do not
dominate the integrated $I_D$ values. The additional exact-boundary maxima on
JBC faces 7 and 38 were separately checked, and their shrinking-neighborhood
contributions also decay toward zero.

The native-BRep positive and negative components reported in
Table~\ref{tab:publicsignatures} retain elliptic and saddle/reverse
developability intensity separately. The corresponding signed all-hull
fields remain available as secondary reproducibility assets, but are
not repeated in the manuscript because the curvature-class field below
preserves the main spatial distinction more compactly.

\paragraph{Curvature composition.}
Curvature-class fractions measure areal composition rather than
curvature intensity. Figure~\ref{fig:classpanels} localizes the flat,
single-curvature, elliptic, and saddle/reverse regions whose
native-BRep represented-valid-area fractions are reported in
Table~\ref{tab:publicsignatures}. Isolated singular sets carry zero
represented area and therefore do not prevent the class fractions from
being evaluated, although their provenance remains recorded.

Intensity and areal coverage are complementary. A localized region of
large double curvature and an extensive region of moderate double
curvature need not produce either the same $I_D$ or the same
curvature-class fractions. The quasi-zero thresholds used for the
classification should therefore be interpreted as numerical
scale definitions rather than calibrated plate-forming categories.

\subsection{Representation sensitivity: native BRep versus triangulated geometry}
\label{sec:results_representation}

The representation study separates two effects that should not be conflated.
First, the controlled same-face tests keep the sampled physical geometry fixed
and isolate the error introduced by mesh curvature recovery and triangle
connectivity. Second, the native-BRep and frozen fine-STL rows in
Table~\ref{tab:publicsignatures} compare the actual campaign representations
and therefore include tessellation, resolution, boundary, and local
mesh-quality effects.

Table~\ref{tab:same_face_verification} reports the controlled same-face errors
defined by Eq.~\eqref{eq:sameface_error}. At the fine $48\times48$ level, the
relative $L_2$ errors in mean curvature range from approximately $0.7\%$ to
$4.3\%$, while the corresponding Gaussian-curvature errors range from
approximately $1.4\%$ to $8.2\%$. Consistent and alternating diagonal patterns
give comparable values. Every $48\times48$ error is below its corresponding
$12\times12$ error, although monotone refinement is not claimed for every
sequence; for example, the DTMB 5415 alternating-diagonal $K$ error changes
from 6.805\% at $24\times24$ to 6.848\% at $48\times48$. Across all $H$/$K$
and connectivity sequences, the $48\times48$-to-$12\times12$ error ratios
range from 0.0820 to 0.5546, with every ratio below unity. These tests quantify
matched-geometry curvature-recovery accuracy, refinement behavior, and
connectivity sensitivity, but do not imply full-hull BRep-to-mesh convergence
of the integrated signature.

The frozen fine-STL realizations in Table~\ref{tab:publicsignatures} show a
larger effect at the integrated-signature level. Relative differences in
$I_D$ with respect to native BRep range from $-16.0\%$ for KVLCC2M to
$+24.3\%$ for JBC, while the signed components and curvature-class fractions
also show material redistribution for several hulls. When source CAD is
available, the native-BRep evaluation is therefore used as the reference
realization for the adopted authoritative BRep; the mesh backend enables
assessment when triangulated geometry is the available representation. These
are representation differences, not isolated curvature-estimator errors, and
the two are not treated as interchangeable numerical descriptions of a unique
derivative-based value.

The frozen coarse/medium/fine campaign provides a complementary
mesh-resolution-sensitivity study of the supplied STL representations.
Across the four hulls, absolute coarse-to-fine changes in $I_D$ range from
$16.7\%$ to $21.9\%$, decreasing to $2.5$--$10.3\%$ from medium to fine. For
the curvature-class vector, the corresponding $d_1$ distances defined in
Eq.~\eqref{eq:class_l1} decrease from $0.036$--$0.736$ to
$0.010$--$0.219$. Because the campaign meshes are representation-specific
triangulations rather than exact parametric tessellations of the authoritative
BRep, this is not a controlled convergence-to-native-BRep study. The results
show why representation, resolution, curvature reconstruction, and valid-area
conventions must accompany derivative-based results.

JBC provides the clearest mesh-quality stress case. Despite the comparatively
poor quality of its campaign tessellations, its controlled same-face
curvature-recovery errors remain within the range observed for the other
hulls. This contrast separates curvature-recovery accuracy from representation
and mesh-quality effects and reinforces the need for representation-aware
realization, refinement evidence, and mesh-quality provenance.

\begin{table*}[!t]
\centering
\caption{Controlled same-face BRep-to-mesh curvature-recovery errors at the
fine $48\times48$ level. Values are $100\varepsilon_H$ and
$100\varepsilon_K$, with $\varepsilon_X$ defined by
Eq.~\eqref{eq:sameface_error}, for consistent and alternating triangle-diagonal
patterns.}
\label{tab:same_face_verification}
\small
\begin{tabular}{lrrrr}
\toprule
& \multicolumn{2}{c}{$H$ error [\%]}
& \multicolumn{2}{c}{$K$ error [\%]} \\
\cmidrule(lr){2-3}
\cmidrule(lr){4-5}
Hull & Consistent & Alternating & Consistent & Alternating \\
\midrule
DTMB 5415 & 1.176 & 1.195 & 6.447 & 6.848 \\
KCS       & 4.314 & 4.336 & 8.211 & 7.873 \\
JBC       & 0.904 & 1.050 & 2.107 & 2.180 \\
KVLCC2M   & 0.715 & 0.842 & 1.405 & 1.384 \\
\bottomrule
\end{tabular}
\end{table*}

Table~\ref{tab:signaturevalidity} summarizes the metric-specific validity and
provenance used to qualify the global signature values.

\begin{table*}[!t]
\centering
\caption{Metric-specific validity and provenance for the retained
public-hull signature. The table qualifies, rather than repeats, the
signature values. Mesh-quality labels are curvature-reliability
advisories. Validity terminology is defined in
Section~\ref{sec:metric_validity}.}
\label{tab:signaturevalidity}
\scriptsize
\begin{tabularx}{\textwidth}{
    llr
    >{\raggedright\arraybackslash}X
    >{\raggedright\arraybackslash}X
    >{\raggedright\arraybackslash}X
}
\toprule
Hull
& Representation
& $A_{\rm valid}/A$ [\%]
& Developability status
& $\mathbf{a}_C$ status
& Numerical evidence \\
\midrule

\multirow[c]{2}{*}{DTMB 5415}
& native BRep
& 100.000
& convergent improper
& finite; measure-zero singular-set caution
& nested audit: excluded-area fraction $5.817\times10^{-9}$;
  final $I_D$ change $0.0512\%$ \\

& frozen fine STL
& 90.683
& finite; representation- and refinement-sensitive
& finite; representation- and refinement-sensitive
& medium--fine $|\Delta I_D|/I_{D,f}=2.5\%$; quality: caution \\

\cmidrule(lr){1-6}

\multirow[c]{2}{*}{KCS}
& native BRep
& 100.000
& finite/converged
& finite/converged
& regular control; bounded-refinement $I_D$ change $0.451\%$ \\

& frozen fine STL
& 93.697
& finite; representation- and refinement-sensitive
& finite; representation- and refinement-sensitive
& medium--fine $|\Delta I_D|/I_{D,f}=10.3\%$; quality: poor \\

\cmidrule(lr){1-6}

\multirow[c]{2}{*}{JBC}
& native BRep
& 100.000
& convergent improper
& finite; measure-zero singular-set caution
& nested audit: excluded-area fraction $3.276\times10^{-12}$;
  final $I_D$ change $0.000670\%$ \\

& frozen fine STL
& 97.350
& finite; representation- and refinement-sensitive
& finite; representation- and refinement-sensitive
& medium--fine $|\Delta I_D|/I_{D,f}=6.47\%$; quality: poor \\

\cmidrule(lr){1-6}

\multirow[c]{2}{*}{KVLCC2M}
& native BRep
& 100.000
& convergent improper
& finite; measure-zero singular-set caution
& nested audit: excluded-area fraction $3.584\times10^{-9}$;
  final $I_D$ change $0.0146\%$ \\

& frozen fine STL
& 95.365
& finite; representation- and refinement-sensitive
& finite; representation- and refinement-sensitive
& medium--fine $|\Delta I_D|/I_{D,f}=3.19\%$; quality: caution \\

\bottomrule
\end{tabularx}
\end{table*}

\section{Discussion}
\label{sec:discussion}

The results establish three complementary points that determine how the
proposed signature should be interpreted in early-stage design. First, the
retained descriptors form an engineering hierarchy of curvature intensity,
sign, and areal extent rather than a universal producibility score. Second,
their numerical interpretation requires metric-specific validity and explicit
representation provenance. Third, the present framework remains a geometric
screening layer whose connection to manufacturing effort requires
process-specific experimental or shipyard validation. These points are
discussed in the following subsections.

\subsection{Engineering hierarchy of the signature}
The results support an ordered interpretation of the signature, not a universal score. Among the descriptors retained in the present framework, total developability provides the most direct geometric connection to plate forming because Gaussian curvature distinguishes bending into a developable shape from deformation requiring in-plane strain or plate subdivision. Its signed components retain whether the double curvature is elliptic or saddle/reverse, while curvature-class fractions add the areal extent of flat, singly curved, and the two double-curved surface types. Intensity, sign, and extent answer different production-related geometric questions and should not be collapsed into uncalibrated weights.

Among the four cases, KCS has approximately 24\% larger developability deviation than DTMB 5415, while double-curved regions cover approximately 72.9\% of the KCS valid surface versus essentially all of the DTMB valid surface. This contrast demonstrates the non-redundancy of intensity and extent: stronger average double curvature may be concentrated over a smaller surface fraction, while lower average intensity may extend across nearly the entire valid surface. The signed components add a third distinction by separating elliptic from saddle/reverse intensity.

\subsection{Validity and representation provenance}
Validity and representation are mandatory parts of the scientific result. Native derivatives can expose a singularity that a finite tessellation replaces by a numerical cutoff; changing mesh resolution or connectivity then changes that cutoff. The observed BRep--mesh differences therefore document numerical realization and provenance rather than invalidate the underlying continuous descriptors.

The DTMB 5415 case makes this distinction explicit. Its represented surface area is finite, and the nested-neighborhood analysis shows that its absolute-Gaussian-curvature developability is a convergent improper integral. A single geometry-level validity flag would therefore be insufficient: different functionals may impose different regularity and integrability requirements on the same geometry. Likewise, a finite value obtained at a particular quadrature depth or mesh resolution cannot be promoted to an established metric value when convergence has not been established. Reporting validity per metric is therefore a scientific requirement rather than merely software bookkeeping.

\subsection{Initial screening layer and validation gap}
The proposed signature is an initial geometric screening layer. It can expose double-curvature intensity and type and areal composition before plate layout, material, thickness, forming route, tolerances, equipment, and yard practice are known. It does not establish that one hull is cheaper, requires fewer man-hours, or should follow a particular forming route. The principal future validation gap is correlation and calibration against plate-scale experiments or shipyard production evidence, including measured forming effort, distortion, rework, and inspection outcomes.

Such evidence is also needed to translate numerical quasi-zero thresholds into process-relevant categories. The present benchmark masters contain construction and continuity choices that may not represent every production CAD model, and the mesh campaign shows that resolution, connectivity, topology, and quality remain part of the numerical evidence.

Once calibrated, individual descriptors or distributed fields may serve as objectives, constraints, surrogate responses, or dimensionality-reduction features in design workflows. Those are downstream uses only. No optimization, surrogate, composite index, or yard-specific production model is demonstrated here. The present contribution is the interpretable signature and its validity/provenance contract.

\section{Conclusions}
\label{sec:conclusions}
A representation-aware framework for geometry-based producibility screening
has been formulated and evaluated for external hull surfaces. The proposed
signature combines total and signed developability deviation,
curvature-class area fractions, distributed surface fields, and
metric-specific validity and representation provenance. Its purpose is to
characterize surface features that are geometrically relevant to hull-form
producibility during early design, before plate layout, material, forming
process, tolerances, and shipyard-specific information are available. It is
therefore not intended as a universal producibility score or as a calibrated
predictor of forming effort, fabrication cost, labor, process selection, or
manufacturing feasibility.

The analytical and semi-analytical controls verify the continuous geometric
definitions and the principal numerical operations used by the native-BRep
and mesh realizations. Controlled same-face BRep-to-mesh comparisons further
quantify the accuracy of the selected discrete curvature reconstruction on
matched geometry, independently of CAD-to-mesh correspondence. 
These tests quantify the accuracy, refinement behavior, and connectivity sensitivity of the selected
mesh curvature reconstruction within the adopted verification protocol, but do
not by themselves establish end-to-end convergence of the complete mesh-based
signature.

Application to DTMB 5415, KCS, JBC, and KVLCC2M shows that the components of
the signature provide complementary information. Total developability
deviation characterizes the average intensity of double curvature, its signed
components distinguish elliptic from saddle/reverse contributions, and the
curvature-class fractions quantify their areal extent. KCS, for example,
exhibits approximately 24\% greater developability deviation than DTMB 5415,
while double-curved regions occupy approximately 72.9\% of its valid surface
versus essentially the entire DTMB valid surface. The distributed fields
identify where these global contributions originate. These results demonstrate
that curvature intensity, sign, and areal composition provide related but
non-redundant geometric information, without implying a ranking of overall
manufacturing difficulty among the hulls.

The comparison between native BRep and frozen fine-STL realizations also shows
that derivative-based descriptors can be materially affected by geometric
representation and mesh resolution. When authoritative CAD geometry is
available, the native-BRep evaluation is therefore used as the reference
realization for the represented geometry, whereas mesh-based results retain
their explicit representation and numerical provenance. The two backends are
thus treated as distinct numerical realizations of the same conceptual
descriptors rather than as interchangeable estimates of a unique
representation-independent value. Higher-regularity surface and section-based
quantities were also investigated but were not included in the recommended
signature because robust general hull-level behavior could not be established
across the considered public CAD representations.

The principal remaining physical-validation step is to relate the proposed
geometric descriptors to independent production-relevant evidence, such as
plate-scale forming experiments, deformation or strain measures, rework,
inspection data, or shipyard production records. Such evidence is required
before descriptor magnitudes or numerical classification thresholds can be
interpreted quantitatively in terms of process-specific forming difficulty,
manufacturing effort, or cost. Until such calibration is available, the
signature should be interpreted as an early-stage geometric screening and
localization framework.

Independently of this process-specific calibration, the normalized global
descriptors and distributed fields provide machine-readable geometric
quantities that can be incorporated into computational design workflows.
Individual signature components may be used as objectives or constraints in
hull-form optimization, as responses in surrogate models, or as features in
design-space exploration and dimensionality-reduction methods. Such downstream
applications are not evaluated in the present proof of concept, but they
represent a primary intended use of the framework once appropriate
problem-specific interpretation and validation are established.

The present contribution therefore establishes the geometry-based signature,
its representation-aware numerical realization, and its metric-specific
validity and provenance contract as a foundation for subsequent computational
design applications. The companion \texttt{HullProd} implementation exposes
the recommended global descriptors, distributed fields, validity information,
and representation provenance through a common interface.



\section*{Acknowledgements}
Andrea Serani is supported by the U.S. Office of Naval Research (ONR) under Grant No. N00014-26-1-2164, as part of the BEAM project (``Bayesian Exploration and Optimization for Hull-form Architecture and Producibility Modeling''), under the administration of Dr. Robert Brizzolara.

\section*{Data and code availability}
The \texttt{HullProd} implementation used in this study is publicly available as
version 1.0.1 through GitHub and the Python Package Index (PyPI). The
archived software release is available on Zenodo at \url{https://doi.org/10.5281/zenodo.22288105}. The versioned implementation,
benchmark configurations, and machine-readable provenance support
reproducibility. Externally sourced hull geometries remain
subject to their respective providers' access and redistribution terms.

\bibliographystyle{unsrt}  
\bibliography{references}

\end{document}